\documentclass[11pt]{article}
\usepackage[pdftex]{graphicx}
\usepackage{hyperref}
\usepackage{amsmath}
\usepackage{graphicx}
\usepackage{amssymb}
\renewcommand{\title}[1]{%
    \bigskip%
    \begin{center}%
    \Large\bf #1%
    \end{center}%
    \vskip .2in}

\renewcommand{\author}[1]{%
    {\begin{center}
    #1
    \end{center}}}
\newcommand{\address}[1]{\vspace{-1.7em}\vspace{0pt}
    {\begin{center}
    \it #1
    \end{center}}}

\begin{document}

\title{\bf{A New Action for Nonrelativistic Particle in Curved Background}}

\author
{
Rabin Banerjee  $\,^{\rm a,b}$,
Pradip Mukherjee $\,^{\rm c,d}$}
\address{$^{\rm a}$ S. N. Bose National Centre 
for Basic Sciences, JD Block, Sector III, Salt Lake City, Kolkata -700 098, India }

\address{$^{\rm c}$Department of Physics, Barasat Government College,Barasat, India}
\address{$^{\rm b}$\tt rabin@bose.res.in}
\address{$^{\rm d}$\tt mukhpradip@gmail.com}
\begin{abstract}
We obtain a new form for the action of a nonrelativistic particle coupled to Newtonian gravity. The result is different from that existing in the literature which, as shown here, is riddled with problems and inconsistencies. The present derivation is based on  the formalism of galilean gauge theory, introduced by us as an alternative method of analysing nonrelativistic symmetries in gravitational background.
\end{abstract}

\noindent{\it{Introduction }}

\smallskip
 Nonrelativistic (NR) symmetries in a gravitational background emerged recently as an attractive tool for investigating a wide range of physical
 phenomena in topics as varied as, but not confined to,  condensed matter physics, hydrodynamics and cosmology. 
  The concept of such symmetries is not new, although its ramifications are.
 Indeed, almost simultaneously with Einstein  a (space-time) covariant  geometrical theory of Newtonian gravitation \cite{MTW} was formulated by Cartan  \cite{Cartan-1923, Cartan-1924}. The corresponding Newton - Cartan (NC) manifold has a degenerate metric structure. There is a  rich literature \cite{Havas, Daut, TrautA, Kuch, EHL} on the subject that covers its myriad aspects. Recent applications require 
coupling of matter fields with gravity which is done through the use of these  Newton Cartan metrics.
A number of different approaches to pursue this problem have appeared  in the recent past \cite{SW, J}, the most popular among these is based on the gauging of 
 (extended) Galilean group algebra \cite{PPP, ABPR}. Just as variants of the approach have been followed in the applications  to condensed matter systems, especially fractional quantum Hall effect \cite{SW},\cite{GA}, \cite{A}, new results of fundamental implications have also been mooted  in  giving an action principle for Newtonian gravity \cite{l}.  However, in some cases the approach is confronted with various inconsistencies  like non canonical transformations of the metric,
 wrong flat limit etc., as discussed in \cite{Sunandan}. The consistency of the foundation of a physical theory is intimately connected with the results of its applications. One would certainly like a compact algorithm which is universally applicable, reproduces the proper transformations of the geometric objects (the metric tensors)  and retrieves Galilean symmetry in the flat limit.
 
 A fundamental example is the formulation of an action for a NR particle that couples with gravity \cite {Kuch, PPP, kluson}. It forms the basis of coupling more and more sophisticated theories ( like superparticles or even (super)strings) with background gravity \cite{B}. The action formalism for the NR particle has been discussed in the literature \cite {Kuch, PPP, kluson}, following the methods outlined above. However the result, as shown here, fails on both counts on not having the appropriate transformation for the Newton Cartan metric and having inconsistencies in its passage to the flat limit. 
 
 The motivation of the present work is to provide a suitable new action for a NR particle in a  gravitational background that is not riddled with any problems or ambiguities. It is formulated on a systematic  algorithm for coupling matter fields with gravity which was introduced by us in a set of papers \cite{BMM1, BMM2, BM4}. Named as galilean gauge theory (GGT), it is based on  gauging  the  NR symmetry of a generic model in flat space. Numerous illustrations \cite {BMM1, BMM2, BM4,BMM3, BM5, milne} demonstrated the robustness of the algorithm and  that it fulfills all the requirements without the addition of any extra or ad-hoc structure. The action found here has the appropriate flat limit and involves coupling with the Newton Cartan metric that has the correct transformations of a metric tensor.

\smallskip

\noindent{\it{A critical look at the action for a nonrelativistic point particle }}

\smallskip

The standard action for a NR point particle in flat space is given by,
\begin{eqnarray}
S = \dfrac{1}{2}m\int \dfrac{dx^{k}}{dt}\dfrac{dx^{k}}{dt} dt
\label{actiont}
\end{eqnarray}
This description, however, is not symmetric in the coordinates and time. The parametric description
is more suitable in the analysis of symmetries where
 both space coordinates and time are functions of a parameter $\lambda$;
\begin{equation}
x^{0} = t = t(\lambda)~,~~~ x^{k}(\lambda)                     
\end{equation}
The action in this description can be easily 
obtained from (\ref{actiont}) as,
\begin{equation}
S = \frac{m}{2} \int\frac{x'^k x'^k}{t'} d\lambda\label{actionl}
\end{equation}
where a prime denotes a differentiation with respect to $\lambda$.

The above action is quasi-invariant under the usual infinitesimal global galilean transformations,
\begin{equation}
x^{\mu} \to x^{\mu} + \xi^{\mu}\label{gtst}
\end{equation}
where $\xi^{0} = - \epsilon ~~,~~ \xi^{k} = \epsilon^{k} + w^{k}_{j} x^{j} - u^{k} t$, since the corresponding lagrangian changes as,
\begin{equation}
\delta L = -\dfrac{d}{d\lambda}\bigg(mu^{k}\dfrac{dx^{k}}{d\lambda}\bigg)
\label{variation}
\end{equation}

 The form of the corresponding lagrangian in a curved background is given in the literature \cite{PPP} as,
\begin{equation}
L= \frac{m}{2\Theta t'}\Big(h_{ij}x'^ix'^j + 2A_ix'^i t' + 2A_0 t't'\Big)
\label{new1}
\end{equation}
where $h_{ij}$ are the spatial metric and $A^\mu$ is a new gauge field.

It is invariant under the local galilean transformations where, contrary to the standard galilean transformations, the parameters in (\ref{gtst}) are now space time dependent. Considering the universal role of time in NR theory, the time translation parameter is taken to be a function of time only $\epsilon(t)$, while the other ones are functions of both space and time,
\begin{equation}
\epsilon \to \epsilon(t) \,\,\, ; \,\,\, \epsilon^k, \,\,\,  w^k_j ,\,\,\,  u^k \to f(x, t)
\label{localparameter}
\end{equation}
 The same quasi invariance (\ref{variation}) (now with a space time dependent boost parameter) is reproduced provided the  
 transformation laws of the newly introduced fields are given by,
\begin{eqnarray}
\nonumber
\delta \Theta &=& \dot\epsilon \Theta\\
\nonumber
\delta h_{ij} &=& -h_{kj}\partial_i \xi^k - h_{ki}\partial_j \xi^k\\
\nonumber
\delta A_0 &=& 2\dot\epsilon A_0 - A_i\partial_t\xi^i -\Theta \partial_t (h_{ij}
u^i x^j)\\
\delta A_i &=& \dot\epsilon A_i - A_j\partial_i\xi^j -h_{ij}\partial_t\xi^j - \Theta \partial_i (h_{kj}u^k x^j)
\label{new2}
\end{eqnarray}

The  lagrangian (\ref{new1}) can be put in a more suggestive form that is frequently used in the literature \cite{PPP, kluson, B},
\begin{equation}
L=\frac{m}{2\tilde\tau_\rho x'^\rho}\tilde h_{\mu\nu}x'^\mu x'^\nu - m\phi\tilde\tau_\rho x'^\rho
\label{new3}
\end{equation}
where,
\begin{equation}
\phi=-\frac{1}{\Theta^2}\Big(A_0 - \frac{1}{2} h^{ij}A_i A_j\Big)
\label{new4}
\end{equation}
and $\tilde h_{\mu\nu}$ and $\tilde\tau_\mu$ are elements of the Newton Cartan geometry parametrised as \cite{PPP},
\begin{equation}
\tilde\tau_\mu = [\Theta, 0, 0, 0]\,\,\, ;\,\,\, \tilde h_{\mu\nu}=
\begin{pmatrix}
h^{rs}A_r A_s & A_j\\
A_i & h_{ij}\end{pmatrix}
\label{matrix}
\end{equation}
The two other elements of the Newton Cartan geometry are given by \cite{PPP},
\begin{equation}
\tilde v^\mu = \frac{1}{\Theta}[1; -h^{ij}A_j]\,\,\,;\,\,\, \tilde h^{\mu\nu}=
\begin{pmatrix}
0&0\\0&h^{ij}
\end{pmatrix}
\label{matrix1}
\end{equation}
One may verify that they satisfy the Newton Cartan algebra,
\begin{equation}
 \tilde h^{\mu\nu} \tilde \tau_\nu = \tilde h_{\mu\nu} \tilde v^\nu =0,\,\,\, \tilde v^\mu\tilde\tau_\mu=1,\,\,\ \tilde h^{\mu\nu}\tilde h_{\nu\rho}=
 \delta^\mu_\rho-\tilde v^\mu \tilde\tau_\rho = P^\mu_\rho
 \label{nc2}\end{equation}
  Two other useful relations that are valid in this parametrisation  are,
 \begin{equation}
\tilde h_{0i}= -\Theta  h_{ij} \tilde v^j \,\,\, ;\,\,\, \tilde h_{00}= \Theta^2 \tilde h_{ij} \tilde v^i \tilde v^j
\label{trick} 
 \end{equation}

We now like to stress that there are certain insurmountable obstacles in interpreting (\ref{new3}) as the lagrangian for a NR particle coupled to gravity. These are:

\begin{enumerate}

\item

 There is a problem in taking the flat limit. In this case the new fields $(A)$  vanish, the field $\Theta$ goes to unity indicating the flow of time and the spatial metric 
$h_{ij}$ goes to the Kroneckar delta. Then the lagrangian  in the form (\ref{new1}) reproduces the familiar parametrised invariant NR particle lagrangian given in (\ref{actionl}). But while this is essential for a proper flat limit, it is not sufficient. The point is that there is an anomaly in the transformation law for $A_0$ (\ref{new2}). While the left side vanishes, the right side does not. There is a non-zero contribution from the last term containing the boosts.

\item 

 While the first term in (\ref{new3}) involves the appropriate Newton Cartan coupling, the second does not. Moreover the problems are further compounded by the fact that the Newton Cartan structure $\tilde h_{\mu\nu}$ appearing there does not even transform as a second rank tensor so that it cannot be regarded as an appropriate Newton Cartan metric, despite satisfying the algebra (\ref{nc2}). Hence the meaning and interpretation of the first term also is unclear. This is explicitly shown below for the $0-i$ component.
 

\end{enumerate}
A  covariant second rank tensor, under the infinitesimal coordinate transformations (\ref{gtst})  should transform as,
\begin{equation}
\delta\tilde h_{\mu\nu}= (\delta^\lambda_\mu- \partial_\mu\xi^\lambda) (\delta^\rho_\nu- \partial_\nu\xi^\rho) \tilde h_{\lambda\rho} - \tilde h_{\mu\nu}
\label{change}
\end{equation}
Taking the $0-i$ component and identifying the various pieces 
with (\ref{matrix}), we obtain the following transformation law for the field $A_i$,
\begin{equation}
\delta A_i = \dot\epsilon A_i - A_j\partial_i\xi^j - h_{ij}\partial_t\xi^j 
\label{new5}
\end{equation}
This does not agree with the given transformation law (\ref{new2}), the difference being the last term proportional to the boosts. In fact the same result is obtained if we repeat the analysis for the other components. This is another manifestation of the boost anomaly mentioned in the above noted first point. Indeed if we drop the boost symmetry, then both problems disappear. The transformation for $A_0$ is consistent with the flat limit just as the Newton Cartan metric has the correct transformations. We will return to this point towards the end of the analysis.

  It is clear that it is not just desirable but also essential to provide a new form of the lagrangian that is free from these shortcomings. To derive our result we first introduce the rudiments of galilean gauge theory on which our analysis is based.
  
  \smallskip

  \noindent{\it{Galilean gauge theory and Newton Cartan geometry}}
  
  \smallskip
  
  Over the last few years we have developed a general formalism which is able to construct the curved space generalisation of a given NR theory in flat space \cite{BMM1, BMM2, BM4}. It has been christened as galilean gauge theory in analogy with its relativistic avatar, the Poincare gauge theory \cite{U, BGM}. We have provided several applications of this formalism \cite{BM4, BMM3, BM5, milne} and also exhibited its connection with Newton Cartan geometry \cite{BMM2}. 
  
  The basic idea is to localise the galilean symmetry following (\ref{localparameter}).  For a given NR theory in flat space, the invariance holds when the parameters are global. Naturally, on making the parameters space time dependent, this invariance would be lost. To recover the invariance after localisation, one has to replace the ordinary derivatives by suitable covariant derivatives. This requires the introduction of new fields denoted by $\Lambda_\nu^\alpha$. The requirement of invariance of the action under local transformations fixes the transformation properties of the new fields as \cite{BMM1, BM4},
\begin{equation}
\delta \Lambda^{a}_{\nu} = -\partial_{\nu} \xi^{\beta}\Lambda^{a}_{\beta} + w^{a}_{b}\Lambda^{b}_{\nu} - u^{a} \Lambda^{0}_{\nu} : \delta\Lambda^0_0=\dot\epsilon\Lambda^0_0 : \Lambda^0_i=0
\label{lambdatrans} 
\end{equation}  
   These results are valid irrespective of the specific NR theory under consideration. We shall prove them in the next section for the particle model taken here. This shows that it is  possible to provide a geometric interpretation to these new fields. In fact from the above equation we see that while the (local) indices $a$ are Lorentz rotated, the (global) indices $\nu$ are coordinate transformed. Thus $\Lambda_\nu^\alpha$ may be interpreted as the inverse vielbein connecting the local and global basis,{\footnote{Indices from the beginning of the alphabet denote the local basis while those from the middle indicate the global basis. Greek indices denote space-time while only space is gven by the Latin ones.}}
 \begin{equation}
\hat{e}_{\mu} = \Lambda_{\mu}^{\alpha}\hat{e}_{\alpha}\label{veilbein}
\end{equation}  
 The vielbein $\Sigma$ is obtained by the inverse of $\Lambda$,  
 \begin{eqnarray}
\Lambda_{\mu}^{\alpha} \Sigma_{\beta}^{\mu} & =& \delta^{\alpha}_{\beta}\nonumber\\
\Lambda_{\mu}{}^{\alpha} \Sigma_{\alpha}{}^{\nu}& =& \delta^{\nu}_{\mu}
\label{inverserelationsw}
\end{eqnarray}
   and is used for inverting (\ref{veilbein}),
 \begin{equation}
\hat{e}_{\alpha} = \Sigma_{\alpha}^{\mu}\hat{e}_{\mu}
\end{equation}

In flat space there is no difference between the global and local basis and the veilbeins simply reduce to the appropriate Kroneckar deltas. It is now possible to give a metric formulation, based on these vielbeins, that reproduces the Newton Cartan geometry.
 We introduce the two degenerate metrics of this geometry, a rank 3 spatial metric $h^{\mu\nu}$ and a temporal one form $\tau_\mu$ \cite{BMM2},
 \begin{equation}
 h^{\mu\nu} = \Sigma^\mu_a \Sigma^\nu_a, \,\,\, \tau_\mu= \Lambda_\mu^0 =\Theta\delta_\mu^0
 \label{degeratematric}\end{equation}
 Two more covariant structures are introduced,
 \begin{equation}
 h_{\mu\nu} = \Lambda_\mu^a \Lambda_\nu^a, \,\,\, v^\mu=\Sigma^\mu_0
 \label{degeratematric1}\end{equation}
 It is simple to check that they satisfy the Newton Cartan algebra (\ref{nc2}) and also the additional relations (\ref{trick}). Furthermore, the various metrics satisfy the proper transformations, as expected for tensors. This has been shown by us earlier in \cite{ BM4, BMM3}. 
 
 One may wonder that our Newton Cartan  metrics have the proper transformations in contrast to  that in  (\ref{matrix}). This is because, despite both the forms for the metrics satisfying the Newton Cartan algebra, the two are different. This is the well known ambiguity where different Newton Cartan metrics are related by the Milne boost symmetry which has a natural explanation in the context of GGT \cite{milne}. There are two versions of this symmetry. Either the forms for $\tau_\mu$ and $h^{\mu\nu}$ are preserved and $v^\mu$ and $h_{\mu\nu}$ modified or it is the other way round. As we can easily see it is the first option that is relevant here. Then the Milne symmetry is manifested by the relations,
\begin{eqnarray}
\tilde v^\mu &=& v^\mu + h^{\mu\nu} \psi_\nu \,\,\, ;\,\,\, \tilde\tau_\mu
=\tau_\mu \\
\tilde h_{\mu\nu} &=& h_{\mu\nu} - (\tau_\mu P^\rho_\nu +\tau_\nu P^\rho_\mu)\psi_\rho
+ \tau_\mu\tau_\nu h^{\rho\sigma} \psi_\rho\psi_\sigma \,\, ; \, \tilde h^{\mu\nu}=h^{\mu\nu}\label{milne}
\end{eqnarray}
where $\psi_\mu$ is an arbitrary vector field and the projection operator $P^\mu_\nu$ is defined in (\ref{nc2}). Using the various structures for the metrics it is possible to show that while $\psi_0$ remains arbitrary, $\psi_i$ is determined as,
\begin{equation}
\psi_i  = -\Big(h_{ij} v^j + \frac{A_i}{\Theta}\Big)
\label{milne1}
\end{equation}
This shows the connection between our forms for the Newton Cartan metrics and that used in (\ref{matrix}) and (\ref{matrix1}).
\smallskip
      
 \noindent{\it{Nonrelativistic particle in curved background from galilean gauge theory}}
 \smallskip
 
 We are now ready to derive the action for a NR particle in curved background using the formulation of GGT. We first see the mechanism of the invariance of the flat space theory (\ref{actionl}) under global galilean symmetry. Explicit use of the variations of the derivatives is necessary. These are given by,
 \begin{equation}
\delta \dfrac{dx^{0}}{d\lambda} = \dfrac{d}{d\lambda}(\delta x^{0}) = - \dfrac{d\epsilon}{d\lambda} = 0
\label{der1}
\end{equation}
as $\epsilon $ is constant and,

\begin{equation}
\delta \dfrac{dx^{k}}{d\lambda} = w^{k}_{j}\dfrac{dx^{j}}{d\lambda} - v^{k}\dfrac{dx^{0}}{d\lambda}
\label{der2}
\end{equation}

The change of the action (\ref{actionl}) is then boundary terms only, given by (\ref{variation}). The same equations of motion follow from both the original and the transformed action. So the theory is invariant under the global galilean transformations.

We will now proceed to localize the galilean symmetry
of the model (\ref{actionl}), applying the algorithm of GGT developed by us. It will be useful to express the derivative as
\begin{eqnarray}
\dfrac{df}{d\lambda} = \dfrac{dx^{\mu}}{d\lambda} \dfrac{\partial f}{\partial x^{\mu}}
\end {eqnarray}
Then,
\begin{equation}
\dfrac{dx^{\mu}}{d\lambda} = \dfrac{dx^{\nu}}{d\lambda} \partial_{\nu}x^{\mu}
\end{equation}
The first step is to make the parameters of the global transformation functions of space and time following the prescription (\ref{localparameter}). 

The second step follows logically from the first. The local transformations (\ref{localparameter}) are structurally same as the global galilean transformations only in the neighborhood of a point. We therefore erect everywhere a local coordinate system. The local basis at this stage, is trivially connected with the global basis
\begin{equation}
\hat{e}^{a} = \delta^{a}_{k}\hat{e}^{k}
\label{flat}
\end{equation}
The transformations with the localized parameters can be called a local Galilean transformation with reference to the local coordinates. If we formulate the theory with respect to the local coordinates, the transformation of the coordinates remain formally the same but the derivatives cease to vary as (\ref{der1}) and (\ref{der2}). The next step of our algorithm is to replace $\frac{dx^{\alpha}}{d\lambda}$ by  $\frac{Dx^{\alpha}}{d\lambda}$, where
\begin{equation}
\dfrac{Dx^{\alpha}}{d\lambda} = \dfrac{dx^{\nu}}{d\lambda} \Lambda^{\beta}_{\nu}\partial_{\beta}x^{\alpha} = \dfrac{dx^{\nu}}{d\lambda} \Lambda^{\alpha}_{\nu}
\label{red}
\end{equation}
Here $\Lambda^{\beta}_{\nu}$ are a set of new compensating (gauge) fields, the transformations of which will ensure
that the 'covariant derivatives' will transform in the same way as the usual derivatives do in the global theory. Then the new theory obtained by replacing the ordinary derivatives by the `covariant derivatives' will be invariant under the local gauge transformations. This is the essence of the gauge principle. It is not difficult to calculate these transformations,
\begin{equation}
\delta \dfrac{Dx^{\alpha}}{d\lambda}  = \dfrac{d\xi^{\nu}}{d\lambda} \Lambda^{\alpha}_{\nu} + \dfrac{dx^{\nu}}{d\lambda} \delta \Lambda^{\alpha}_{\nu}
\nonumber
\end{equation}
Taking $\alpha = 0$

\begin{equation}
\delta \dfrac{Dx^{0}}{d\lambda}  = (\delta \Lambda^{0}_{0} - \dot{\epsilon} \Lambda^{0}_{0})\dfrac{dx^{0}}{d\lambda}+ \dfrac{d\xi^{i}}{d\lambda}\Lambda_i^0 ++  \dfrac{dx^{i}}{d\lambda}\delta\Lambda_i^0 
\end{equation}

To keep it covariant i.e. formally the same as (\ref{der1}), we require the above expression to vanish, so that,
\begin{equation}
\delta \Lambda^{0}_{0} = \dot{\epsilon} \Lambda^{0}_{0}\,\,\, , \Lambda_i^0=0
\label{evar1}
\end{equation}
Likewise, for $\alpha = a$

\begin{equation}
\delta \dfrac{Dx^{a}}{d\lambda}  = (\delta \Lambda^{a}_{l} + \partial_{l} \xi^{k}\Lambda^{a}_{k})\dfrac{dx^{l}}{d\lambda} + \dfrac{dx^{0}}{d\lambda} (\delta \Lambda^{a}_{0} - \dot{\epsilon} \Lambda^{a}_{0}+ \partial_0\xi^k\Lambda_k^a)
\nonumber
\end{equation}
According to our requirement, we should get the covariant transformation law that formally looks like (\ref{der2}), replacing the ordinary derivatives by the covariant ones,
\begin{equation}
\delta \dfrac{Dx^{a}}{d\lambda} = w^{a}_{b}\dfrac{Dx^{b}}{d\lambda} - u^{a}\dfrac{Dx^{0}}{d\lambda}
\nonumber
\end{equation}
Solving the above equation yields,
\begin{equation}
\delta \Lambda^{a}_{l} + \partial_{l} \xi^{k}\Lambda^{a}_{k} - w^{a}_{b} \Lambda^{b}_{l} = 0\label{pradip}
\end{equation}
and,
\begin{equation}
\delta \Lambda^{a}_{0} - \dot{\epsilon} \Lambda^{a}_{0} + u^{a} \Lambda^{0}_{0} + \partial_{0} \xi^{k}\Lambda^{a}_{k} - w^{a}_{b}\Lambda^{b}_{0} = 0
\label{evar2}
\end{equation}
The relations (\ref{evar1}), (\ref{pradip}) and (\ref{evar2}) reproduce the desired result (\ref{lambdatrans}), as announced there. 

We have thus proved that if the new fields obey (\ref{lambdatrans}),  then 
$\frac{Dx^{\alpha}}{d\lambda}$ transforms under local galilean transformations  in the same way as
$\frac{dx^\mu}{d\lambda}$ transforms under the global Galilean transformations. Hence the action,
\begin{equation}
S = 
\int
\dfrac{1}{2}m\dfrac{ \dfrac{Dx^{a}}{d\lambda}\dfrac{Dx^{a}}{d\lambda}}{\bigg(\dfrac{Dx^{0}}{d\lambda}\bigg)} ~d\lambda
\label{actionm}
\end{equation}
obtained from (\ref{actionl}) by substituting $\frac{dx^\mu}{d\lambda}$ by $\frac{Dx^{\alpha}}{d\lambda}$ in (\ref{actionl}) is the desired action. In fact, substituting  (\ref{lambdatrans}) in the variation of (\ref{actionm}) we can compute
\begin{equation}
\delta S = 
-\int
 m u^{a} \dfrac{Dx^{a}}{d\lambda} ~d\lambda
 \label{lvar}
\end{equation}
It is formally the same as (\ref{variation}). In case of the global Galilean transformations, invariance of the action followed immediately since $u^a$ was a constant parameter. Here we have to show that the integrand is indeed a total derivative.

Using Leibnitz rule, which trivially follows from the definition of the covariant derivatives,
\begin{eqnarray}
m u^{a} \dfrac{Dx^{a}}{d\lambda} = \dfrac{D(mu^{a}x^{a})}{d\lambda} - mx^{a}\dfrac{Du^{a}}{d\lambda}
\end{eqnarray}

Now, from (\ref{red}),
\begin{eqnarray}
\dfrac{Du^{a}}{d\lambda} = \dfrac{dx^{\nu}}{d\lambda} \Lambda^{\alpha}_{\nu} \partial_{\alpha}u^{a}
\end{eqnarray}
When we go to the global Galilean transformation limit,
$\Lambda^{\alpha}_{\nu} \to \delta^{\alpha}_{\nu}$ and  $\dfrac{Du^{a}}{d\lambda} \to \frac{du^a}{d\lambda}$. 
Then, isolating the spacetime dependence of $\Lambda_\mu^\alpha$ as a first order correction,
\begin{equation}
\Lambda_\mu^\alpha = \delta_\mu^\alpha + \epsilon_\mu^\alpha (x, t)
\label{correction}
\end{equation}
we obtain,
\begin{equation}
\frac{Du^a}{d\lambda}= \frac{du^a}{d\lambda} + \frac{dx^\nu}{d\lambda}\epsilon_\nu^\alpha\partial_\alpha u^a
\end{equation}
The second term on the right side is quadratic in the infinitesimals and hence is dropped.
 Then upto first order,
\begin{eqnarray}
\dfrac{Du^{a}}{d\lambda} = \dfrac{du^{a}}{d\lambda} = 0
\end{eqnarray}
Again, upto the same order,
\begin{eqnarray}
\dfrac{D(u^{a}x^{a})}{d\lambda} = \dfrac{d(u^{a}x^{a})}{d\lambda}
\end{eqnarray}
Inserting all these in (\ref{lvar}) we get
\begin{equation}
\delta S = 
-\int d( m x^{a}u^{a})
\label{final1}
\end{equation}
The modified action (\ref{actionm}) is thus invariant under local  Galilean transformations, as it should be according to the general algorithm of GGT developed in \cite{BMM1, BMM2}.

 It is now possible to express the action (\ref{actionm}) in a covariant form using the elements of Newton-Cartan geometry. From 
(\ref{red}) and (\ref{degeratematric1}) we get,
\begin{equation}
\dfrac{Dx^{a}}{d\lambda}\dfrac{Dx^{a}}{d\lambda} = h_{\nu\sigma}\dfrac{dx^{\nu}}{d\lambda}  \dfrac{dx^{\sigma}}{d\lambda} 
\end{equation}
so that the final form looks like, 
\begin{equation}
S = 
\int  \left(\dfrac{m}{2\Theta}\right)h_{\nu\rho}\frac{{x'^\nu} x'^{\rho}}{x'^0} ~d\lambda
\label{actionm1}
\end{equation}
where we have set $\Lambda_0^0=\Theta$. This is the same $\Theta$ that appeared earlier in (\ref{new1}). This identification is done on the basis of identical transformations (\ref{new2}) and (\ref{evar1}).  In terms of the usual time variable the action has the form,
\begin{equation}
S = 
\int
\Big(\dfrac{m}{2\Theta}\Big) h_{\nu\rho}\dfrac{dx^{\nu}}{dt}\dfrac{dx^{\rho}}{dt} ~dt
\label{actionm2}
\end{equation}

Expressions (\ref{actionm1}) and (\ref{actionm2}) are the cherished forms for the action of a NR particle coupled to gravity. Introducing the invariant measure by a scaling,
\begin{equation}
dT=\Theta dt
\label{invneasure}
\end{equation}
that satisfies,
\begin{equation}
\delta (dT) =0
\label{vanish}
\end{equation}
which  follows from (\ref{gtst})and (\ref{evar1}), it is possible to rewrite (\ref{actionm2}) in a manifestly space-time covariant form,
\begin{equation}
S = 
\int
\Big(\dfrac{m}{2}\Big) h_{\nu\rho}\dfrac{dx^{\nu}}{dT}\dfrac{dx^{\rho}}{dT} ~dT
\label{actionm3}
\end{equation}

The space-time covariant forms of the action presented either in the parametrised version involving $\lambda$ or the usual time variable are new in the literature. It is also important to mention that the metric formulation of the theory given in (\ref{actionm3}) involves $h_{\mu\nu}$ which is an element of the Newton-Cartan structure. This shows that the coupling is indeed with NR (Newtonian) gravity.

The two shortcomings of the lagrangian (\ref{new3}) do not occur here. The flat limit poses no problems. In this limit we recall that the vielbeins reduce to the Kroneckar deltas. Then $\Theta =1$ which implies $T=t$, $h_{00}=h_{0i}=0$ and $h_{ij}=\delta_{ij}$. The action (\ref{actionm3}) reduces to the standard NR action for a free partiacle in flat space. Contrary to (\ref{new3}) there are no gauge field terms and we do not have to worry about the consistency of their transformations. Also, as mentioned earlier, the Newton Cartan metric $h_{\mu\nu}$ correctly transforms as a second rank covariant tensor.

Before closing the section we mention that if boosts are ignored then  the lagrangian given in the literature \cite{Kuch, PPP} agrees with our result. This is most easily seen by taking the form (\ref{new1}) and comparing with (\ref{actionm1}). In this restricted symmetry case we may identify,
\begin{equation}
2 A_0 = h_{00} \,\,\, ;\,\,\, A_i = h_{0i}
\label{last}
\end{equation}
which is proved by the equality of the transformations on either side where $h_{\mu\nu}$ is defined in (\ref{degeratematric1}). Then our result (\ref{actionm1}) is reproduced. That boosts play a spoilsport  may also be realised from the fact that while it is possible to identify the vielbeins $\Lambda^0_0$ and $\Lambda_i^j$ of our approach with corresponding fields $\Theta$ and $E_i^j$ in \cite{PPP}, there is no analogue for $\Lambda_0^i$. It is simply nonexistent.

\smallskip

\noindent{\it{Final comments}}

\smallskip

We have successfully constructed the action for a NR particle coupled to Newtonian gravity following the systematic procedure that was used for formulating galilean gauge theory (GGT) \cite{BMM1, BMM2, BM4}. The properties of the vielbeins found in GGT were independently derived here for the particle case. Until now all our applicatione \cite{BM4,BM5,milne} were confined to field theoretical models. Its success for the particle case further cements the universality of the approach. 

No ad-hoc and/or arbitrary introduction of gauge field terms was done. Indeed there is no need for this additional baggage that leads to all sorts of complications and inconsistencies.  There is no compelling justifiable reason for their presence, as is mandatory in other approaches. But the clinching point is that, inspite of these new fields, one did not get a consistent theory, as we have elaborated. The final metric formulation given by us in (\ref{actionm3}), on the other hand, just involves the barest essentials- the four vectors and the Newton Cartan metric.

The present work will have applications in various contexts  The covariantisation of the NR particle model served as a bedrock for the covariantisation of other more sophisticated theories like the NR spinning particle model \cite{spin} or superparticle 
and (super)strings in a Newton Cartan background \cite{B}. Even cosmological implications for NR gravity consider, as a starting point, analysis with Newton Hooke gravity where one introduces a cosmological term in Newtonian gravity \cite{newtonhooke}. In light of the present paper, there could appear many surprises in these areas. We hope to return to some of these issues in a future work
.


\begin{thebibliography}{999}
 
 \bibitem{MTW}C. Misner, K. Tnorn and J. Wheeler,Gravitation (Freeman, 1973)

  \bibitem{Cartan-1923}
E.~Cartan, {\it {Sur les vari\'et\'es \`a connexion affine et la th\'eorie de
  la relativit\'e g\'en\'eralis\'ee. (premi\`ere partie)}},  {Annales
  Sci.Ecole Norm.Sup.} {\bf 40} (1923) 325--412.

\bibitem{Cartan-1924}
E.~Cartan, {\it {Sur les vari\'et\'es \`a connexion affine et la th\'eorie de
  la relativit\'e g\'en\'eralis\'ee. (premi\`ere partie) (Suite).}},  {
  Annales Sci.Ecole Norm.Sup.} {\bf 41} (1924) 1--25.
  

   \bibitem{Havas}     
    P. Havas,
Rev. Mod. Phys. {\bf{36}},(1964),938.

\bibitem{Daut}       
    G. Dautcourt:
    {\rm ``Die Newtonske Gravitationstheorie als Strenger Grenzfall
         der Allgemeinen Relativit\"atheorie''},
    Acta Phys. Pol. {\bf{25}},5,(1964),637.



\bibitem{TrautA}     
    A. Trautman, ``Theories of Space, Time and Gravitation" 
    in {\it Lectures on General Relativity},
    S. Deser and K.W. Ford, eds., Prentice-hall, Englewood Cliffs, 1965.
    
  


\bibitem{Kuch}           
   K. Kucha\v{r},
   Phys. Rev. ,{\bf{22D}},6,(1980),1285.




\bibitem{EHL}       
    J. Ehlers:
{\rm ``On Limit Relations between, and Approximative Explanations of,
Physical Theories''},
in {\it Logic, Methodology and Philosophy of Science}, VII,
B. Marcus et al., eds., Elsiever, Amsterdam (1986),405.

 \bibitem{SW}
 D.T. Son and M. Wingate,
Annals.\ of.\ Physics.\ {\bf 321}, {197-224} (2006). 



\bibitem{J}
K.~Jensen,
  SciPost Phys.\  {\bf 5}, no. 1, 011 (2018)
  

 

\bibitem{PPP}
  R.~De Pietri, L.~Lusanna and M.~Pauri,
  Class.\ Quant.\ Grav.\  {\bf 12}, 219 (1995)
\bibitem{ABPR} R. Andringa, E.~A. Bergshoeff, S. Panda and M. de Roo,
  Class.\ Quantum Grav.\  {\bf 28} 105011 (2011).
 

 
\bibitem{GA} 
  A.~Gromov and A.~G.~Abanov,
  Phys.\ Rev.\ Lett.\  {\bf 113}, 266802 (2014)
  doi:10.1103/PhysRevLett.113.266802
  
\bibitem{A} 
  R.~Auzzi, S.~Baiguera and G.~Nardelli,
  Phys.\ Rev.\ D {\bf 97}, no. 8, 085010 (2018)
  
\bibitem{l} 
  D.~Hansen, J.~Hartong and N.~A.~Obers,
  Phys.\ Rev.\ Lett.\  {\bf 122}, no. 6, 061106 (2019)
  
  \bibitem{Sunandan}
  R.~Banerjee, S. Gangopadhyay and P.~Mukherjee, Int.J.Mod.Phys. A32 (2017) , 1750115 
  
  \bibitem{kluson} J. Kluson, Eur. Phys. Jour. C78 (2018) 117
  
  \bibitem{B} E. Bergshoeff, J. Rosseel and T. Zojer, Class. Quant. Grav. 32 (2015) 205003
  

 

  \bibitem{BMM1} 
  R.~Banerjee, A.~Mitra and P.~Mukherjee,
  Phys.\ Lett.\ B {\bf 737}, 369 (2014).

  \bibitem{BMM2}R. Banerjee, A. Mitra, P. Mukherjee, Class.\ Quantum Grav.\ {\bf{32}}, 045010 (2015).
  
  \bibitem{BM4}R. Banerjee, P. Mukherjee, Phys. Rev. {\bf{D93}}, 085020 (2016)
  
  \bibitem{BMM3}R. Banerjee, A. Mitra, P. Mukherjee, Phys. Rev. D 91, 084021 (2015).

  
  .
 
 \bibitem{BM5} 
  R.~Banerjee and P.~Mukherjee,
   Class.\ Quant.\ Grav.\  {\bf 33}, no. 22, 225013 (2016)


 
\bibitem{milne} R. Banerjee and P. Mukherjee, Phys. Lett. B778, (2018), 303. 

\bibitem{U}  R.~Utiyama,
    Phys.\ Rev.\  {\bf 101} (1956) 1597.
 
   \bibitem {BGM}M. Blagojevic and F.W. Hehl (eds.), ``Gauge Theories of G
ravitation'',
 Imperial College Press, London (2013).
 
 \bibitem{spin} A. Barducci, R. Casalbouni and J. Gomis, JHEP 1801 (2018) 002
 
 \bibitem{newtonhooke} Y. Tian, H.Y. Guo, C-G Huang, Z. Xu, and B. Zhou, Phys. Rev. D71 (2005) 04430







\end{thebibliography}
\end{document}